\documentclass[aps,pre,twocolumn,longbibliography,superscriptaddress]{revtex4}
\usepackage[latin1]{inputenc}
\usepackage[english]{babel}
\usepackage{bm,graphicx,graphics,amsmath,amssymb,epsfig,color}
\usepackage{dcolumn}

\newcommand{\mA}{\mathcal{A}}

\newcommand{\mP}{\mathcal{P}}

\newcommand{\mL}{\mathcal{L}}

\newcommand{\mZ}{\mathcal{Z}}

\newcommand{\be}{\begin{equation}}
\newcommand{\ee}{\end{equation}}
\newcommand{\bea}{\begin{eqnarray}}
\newcommand{\eea}{\end{eqnarray}}
\newcommand{\bse}{\begin{subequations}}
\newcommand{\ese}{\end{subequations}}
\newcommand{\comment}[1]{}

\newcommand{\ggcol}[1]{{\color{black} #1}}

\begin{document}

\title{Symplectic quantization III: Non-relativistic limit}

\author{Giacomo Gradenigo}
\affiliation{Gran Sasso Science Institute, Viale F. Crispi 7, 67100 
  L'Aquila, Italy}
\email{giacomo.gradenigo@gssi.it}
\affiliation{INFN-LNGS, Via G. Acitelli 22, 67100 Assergi (AQ), Italy}

\author{Roberto Livi} \affiliation{Dipartimento di Fisica e Astronomia
and CSDC, Universit\`a di Firenze, via G. Sansone 1 I-50019, Sesto
Fiorentino, Italy}
\email{roberto.livi@unifi.it}
\affiliation{INFN, Sezione di Firenze, via G. Sansone 1, 
I-50019, Sesto Fiorentino, Italy} 
\affiliation{CNR-ISC, via Madonna del Piano 10, 
I-50019 Sesto Fiorentino, Italy}

\author{Luca Salasnich} \affiliation{Dipartimento di Fisica e
Astronomia ``Galileo Galilei'' and Padua QTech Center, Universit\`a di
Padova, via F. Marzolo 8, I-35131 Padova, Italy}
\email{luca.salasnich@unipd.it}
\affiliation{INFN, Sezione di Padova, via F. Marzolo 8,  
I-35131, Padova, Italy} 
\affiliation{CNR-INO, via N. Carrara 1, 
I-50019 Sesto Fiorentino (Firenze), Italy}

\begin{abstract}
First of all we shortly illustrate how the symplectic quantization
scheme [Gradenigo and Livi, 2021] can be applied to a relativistic
field theory with self-interaction. Taking inspiration from the
stochastic quantization method by Parisi and Wu, this procedure is
based on considering explicitly the role of an intrinsic time
variable, associated with quantum fluctuations.  The major part of
this paper is devoted to showing how the symplectic quantization
scheme can be extended to the non-relativistic limit for a
Schr\"odinger-like field.  Then we also discuss how one can obtain
from this non-relativistic theory a linear Schr\"odinger equation for
the single-particle wavefunction. This further passage is based on a
suitable coarse-graining procedure, when self-interaction terms can be
neglected, with respect to interactions with any external field.  In
the Appendix we complete our survey on symplectic quantization by
discussing how this scheme applies to a non-relativistic particle
under the action of a generic external potential.
\end{abstract}

	
\maketitle

\section{Introduction}
\label{intro}

Symplectic quantization has been recently introduced \cite{GL21,G21}
as a quantization scheme suitable for quantum-relativistic field
theories.  It amounts to a deterministic version of the stochastic
quantization method proposed several years ago by Parisi and Wu
\cite{PW81}, and later adopted by several authors
\cite{CR82,DFF83,G83,DH88,KN96}. It is worth recalling that stochastic
quantization aims at computing field correlators in the limit $\tau
\to \infty$, where $\tau$ is interpreted as a {\sl fictitious} time
appearing in a Langevin-like equation. In this sense, there is no need
to attribute to $\tau$ any physical interpretation, relying on the
assumption that the stochastic evolution eventually samples a suitable
measure in configuration space, epitomizing the statistical effect of
quantum fluctuations. In fact, it was also argued that stochastic
quantization amounts to a functional formulation of quantum field
theory, that is formally equivalent to the canonical partition
function of equilibrium statistical mechanics, provided a Wick
rotation to an imaginary coordinate-time is adopted \cite{DFF83}.

Conversely, the basic idea of symplectic quantization is that the
statistical effect of quantum fluctuations can be sampled by a
deterministic symplectic dynamics, driven by an {\sl intrinsic} time
$\tau$, different from the coordinate time of quantum relativistic
field theories. Accordingly, the main outcome of this approach is the
possibility of looking at the dynamics of a quantum-relativistic
field-theory as a relavant piece of physical information, while the
system approaches asymptotically equilibrium. In this respect it is
important to point out the basic hypothesis of symplectic
quantization: the deterministic evolution equations of any
quantum-relativistic field, ruled by the {\sl intrinsic} time $\tau$
is {\sl ergodic}, i.e. it samples a constant-action hyper-surface
according to a uniform probability measure. Such an assumption stems
from the consideration that quantum-relativistic fields are infinite
dimensional functions and there is no a priori reason to be invoked
contrary to ergodicity, once all symmetries of the model under
scrutiny have been taken into account.  The main ingredients of this
approach are summarized in Section \ref{sec:1}, in order to provide a
preliminary framework for the main content of this paper.  In fact,
the main purpose of this paper is to show how the symplectic
quantization procedure applies also to non-relativistic field
theories, while recovering also the limit of standard quantum
mechanics.

In Section \ref{sec:2} we illustrate how symplectic quantization can
be extended to non-relativistic field theories by performing a
suitable non-relativistic limit of the symplectic evolution equations,
which transform into symplectic evolution equations for a
Schr\"odinger field, which still keeps a dependence on the intrinsic
time $\tau$. We discuss also the connection of this non-relativistic
scheme with the standard formulation of non-relativistic QFT: the
Feynman path-integral turns out to be the Fourier-transform of the
partition volume of the constant-action hypersurface $\Omega({\mathcal
  A})$. This outcome elucidates also that symplectic quantization is
the result of a deterministic dynamical evolution yielding a
generalized microcanonical measure, whose entropy is given by
\begin{align}
S_{\text{sym}} = \log (\Omega({\mathcal A})).
\end{align}
In Section \ref{sec:3} we go through the symplectic quantization
scheme for a quantum field in interaction with a classical
electromagnetic field and we show that, by a coarse-graining procedure
over the fast dynamics ruled by $\tau$ (a sort of {\sl adiabatic
  elimination} procedure) one can recover the standard quantum
mechanical limit in terms of the Schr\"odinger equation, provided
self-interaction terms can be neglected with respect to the
interaction with external classical fields. \ggcol{In this perspective
  the symplectic quantization approach to non-relativistic field
  theories might be interpreted as a sort of formulation of quantum
  mechanics relying upon a hidden-variable theory. These are the
  momenta conjugated to fields yielding the dynamics engendered by the
  intrinsic time $\tau$ of quantum fluctuations. Note that this is not
  contradictory to the foundations of quantum mechanics and, in
  particular, with Bell's theorem.  Actually, this does not rule out
  {\sl non-local} theories with hidden variables (see, for instance
  ~\cite{R77}), like those characterized by global conservation
  constraints, as in the case of symplectic quantization.}  In Section
\ref{sec:3a} we report conclusions and perspectives of the matter
contained in this paper. Finally, in the Appendix we provide further
support to the generality of the symplectic approach, by reconsidering
its interest also for ``classical'' quantum models.

\section{Relativistic formulation}
\label{sec:1}

Let us start from the Lagrangian density of a massive charged scalar
field $\varphi(x)$, where $x=x^{\mu}=(x^0,{\bf x})$ with $x^0=c~t$,
$c$ the speed of light in vacuum, $t$ the time, and ${\bf x}\in
\mathbb{R}^3$ the vector of spatial coordinates.  Including the
quartic non-linear self-interaction the Lagrangian density reads
\begin{eqnarray}
\label{eq:lagrangian}
\mathcal{L}(\varphi,\partial_\mu\varphi)  &=& \partial^\mu \varphi^* 
\partial_\mu\varphi - \frac{m^2 \, c^2}{\hbar^2} \, |\varphi|^2 - 
\frac{\lambda}{2} |\varphi|^4
\\
&=& {1\over c^2} |\partial_t \varphi|^2 - |\partial_{\bf x} \varphi|^2 
 - \frac{m^2 \, c^2}{\hbar^2} \, |\varphi|^2 - 
\frac{\lambda}{2} |\varphi|^4 \; . 
\nonumber
\end{eqnarray}
Since in what follows we want to consider the non-relativistic limit
of the theory, we have to consider a complex, i.e. {\sl charged},
field $\varphi$. The integral over the space-time continuum of the
Lagrangian density in Eq.~(\ref{eq:lagrangian}) is the
relativistic-invariant action
\begin{equation}
S[\varphi(x)] = \int \mathcal{L}(\varphi,\partial_\mu\varphi) \, d^4x
\, , 
\label{action}
\end{equation}
where $d^4x=dx^0 \ d^3{\bf x}=c \ dt \ d^3{\bf x}$. 

\section{Symplectic quantization}

The fundamental idea of symplectic quantization is that $x^0=c~t$ is
just a standard coordinate and that one has to consider an additional
time variable, i.e. the {\sl intrinsic time} $\tau$ parametrizing the
sequence of quantum fluctuations~\cite{GL21,G21}. In practice, one
assumes that the complex field depends also on this intrinsic time,
i.e. $\varphi(x;\tau)=\varphi(x^0,{\bf x};\tau)$. Moreover, one
postulates the existence of a generalized ``Lagrangian'' (which has
however the units of an action) containing a sort of {\it kinetic
  energy term}:
\begin{align}
\mathbb{L}[\varphi,\partial_{\tau}{\varphi}] = \int 
c_s^{-2} |\partial_{\tau}{\varphi}(x;\tau)|^2 \, d^4x + S[\varphi(x;\tau)] \; ,
\label{eq:rel-Lagrangian}
\end{align}
where $\partial_{\tau}$ denotes the derivative with respect to $\tau$
and $c_s$ is a suitable parameter with the dimensions of a velocity.
\ggcol{Let us observe that the simple form of the generalized
  Lagrangian in Eq.~\eqref{eq:rel-Lagrangian} is closely inspired,
  almost dictated, by the analogous deterministic dynamics approach to
  Euclidean lattice field theory (see for instance~\cite{CR82}), where
  the presence of the quadratic term in $\partial_\tau\varphi$ is a
  necessary condition to guarantee the sampling of the equilibrium
  measure $\exp(-S_E/\hbar)$.  The new physical constant $c_s$ is
  associated with the intrinsic time-scale of the symplectic dynamics,
  which might be relevant for analyzing transient regimes, but which
  has no practical influence on the sampling of the equilibrium
  distribution of fields.}  Since there is in general no need for
$c_s$ to coincide with $c$, we will keep explicit the dependence on
it. Moreover, in order to illustrate properly how to perform the
non-relativistic limit in the symplectic quantization scheme, it is
convenient to keep explicit the dependence also on the other physical
constants $\hbar$ and $c$.  Following a standard procedure, one can
introduce the canonically conjugated momenta
\begin{align}
\pi(x;\tau) &= \frac{\delta\mathbb{L}}{\delta 
\partial_{\tau}{\varphi}^*(x;\tau)} = 
c_s^{-2} \partial_{\tau}{\varphi}(x;\tau) 
\\
\pi^*(x;\tau) &= \frac{\delta\mathbb{L}}{\delta 
\partial_{\tau}{\varphi}(x;\tau)} = 
c_s^{-2} \partial_{\tau}{\varphi}^*(x;\tau) 
\label{eq:conj-mom}
\end{align}
and thus define the symplectic action, or generalized ``Hamiltonian'', as follows
\begin{align}
& \mathbb{H}[\pi,\varphi] = \nonumber \\
& \int \left[ \pi^*(x;\tau) \partial_{\tau}{\varphi}(x;\tau) + 
\pi(x;\tau) \partial_{\tau}{\varphi}^*(x;\tau)\right] \, d^4x 
- \mathbb{L}[\varphi,\partial_{\tau}{\varphi}] \; . 
\end{align}
Making use of all of the previous equations one can finally write
\begin{align}
\mathbb{H}[\pi,\varphi] = \int 
{|\pi(x;\tau)|^2\over c_s^{-2}} \, d^4x - S[\varphi(x;\tau)] \; .
\label{eq:Ham-symp-rel}
\end{align}

The symplectic dynamical equations then read
\begin{align}
\partial_{\tau}{\varphi}(x;\tau) &= \frac{\delta \mathbb{H}[\pi,\varphi]}
{\delta \pi^*(x;\tau)} \; , 
\\
\partial_{\tau}{\pi}(x;\tau) &= - \frac{\delta \mathbb{H}[\pi,\varphi]}
{\delta \varphi^*(x;\tau)} \; .
\label{eq:sympl-dyn}
\end{align} 
For the field-theory (\ref{eq:lagrangian}) they have the explicit expressions
\begin{eqnarray}
\partial_{\tau}{\varphi}(x;\tau) &=& {\pi(x;\tau)\over c_s^{-2}} \; , 
\\
\partial_{\tau}{\pi}(x;\tau) &=& 
- \left( \Box + {m^2c^2\over \hbar^2} + \lambda |\varphi(x;\tau)|^2\right) 
\varphi(x;\tau), \nonumber \\ 
\label{eq:sympl-dyn-2}
\end{eqnarray}
where $\Box =c^{-2}\partial_t^2 - \partial_{\bf x}^2$. The purpose of
the present discussion has been to introduce the formalism of
symplectic quantization for relativistic fields in order to both
connect with the previous papers on the subject~\cite{GL21,G21} and
also to show how the approach to non-relativistic QFT can be directly
deduced from the relativistic one, which needed therefore to be
presented first.\\

\section{Non-relativistic limit}
\label{sec:2}

Here we discuss the non-relativistic limit in the framework of the symplectic quantization. 
The first step amounts to factorize
the high-frequency component of the field $\varphi(x;\tau)$, thus epitomizing 
its dependence on the parameter $c$ (speed of light):
\begin{align}
\varphi(x;\tau) = {\hbar\over \sqrt{2m c}} \ 
e^{-i \, \frac{m c^2}{\hbar} \, t} \ \Psi({\bf x},t;\tau) \, , 
\label{eq:slow-field}
\end{align}
where $\Psi({\bf x},t;\tau)$ is the non-relativistic field, i.e. the
Schr\"odinger field, which still exhibits a dependence on the
intrinsic time $\tau$. \ggcol{The adiabatic separation in slow and
  fast oscillating components of the field $\varphi(x;\tau)$ is a
  standard approach to perform $1/c$ expansions, see for
  instance~\cite{LL82}.} Let us notice that within the symplectic
quantization approach the distinction between {\it ``classical''} and
{\it ``quantum''} fields is straightforward: anytime a field has
quantum fluctuations it carries an explicit dependence on the
intrinsic time $\tau$, while a classical field does not exhibit any
dependence on $\tau$.\\

By substituting Eq.~(\ref{eq:slow-field}) into the relativistic
action, Eq.~\eqref{action}, and performing the non-relativistic limit
$c\to\infty$ one can conclude that the contribution $\frac{1}{2 c^2}
\Psi^* \frac{\partial^2 \Psi}{\partial t^2} \to 0$ can be neglected,
thus yielding the following expression for the non-relativistic action
\begin{align}
  & S_{\textrm{nr}}[\Psi({\bf x},t;\tau)]  = \int dt \, d^3{\bf x}~ 
\mL_{\textrm{nr}}[\Psi({\bf x},t;\tau)] \nonumber \\
  & \mL_{\textrm{nr}}[\Psi] = 
  {i \hbar\over 2}   
\left( \Psi^* \partial_t \Psi - \Psi \partial_t \Psi^* \right) 
- \frac{\hbar^2}{2m} |\partial_{\bf x} \Psi|^2 -
\frac{g}{2} |\Psi|^4, \nonumber \\
\label{eq:action2}
\end{align}
where $g=\lambda \hbar^4/(4m^2c)$. Then, for consistency, we also
define the {\it slow} non-relativistic component $\Pi({\bf x},t;\tau)$
of the conjugated momentum field $\pi(x;\tau)$ as follows
\begin{align}
\pi(x;\tau) = {\hbar\over \sqrt{2m c}} \ 
e^{-i \, \frac{m c^2}{\hbar} \, t}~\Pi({\bf x},t;\tau) \; . 
\end{align}
In this way we obtain the expression of the generalized non-relativistic Hamiltonian 
\begin{align}
  & \mathbb{H}_{\text{nr}}[\Pi({\bf x},t;\tau),\Psi({\bf r},t;\tau)] = \nonumber \\
  & \int {|\Pi({\bf x},t;\tau)|^2\over \mu} dt~d^3{\bf x} - S_{\textrm{nr}}[\Psi({\bf x},t;\tau)], 
\label{eq:nr-H}
\end{align}
where $\mu=2mc_s^{-2}/\hbar^2$. One can easily recognize in the previous formula 
a generalized {\it separable} Hamiltonian, containing a
generalized ``kinetic energy'' term and a generalized ``potential
energy'' term
\begin{align}
\mathbb{H}_{\text{nr}}[\Pi,\Psi]  = \mathbb{K}[\Pi] + \mathbb{V}[\Psi], 
\label{eq:ham-T-V-1}
\end{align}
with
\begin{align}
  \mathbb{K}[\Pi] & = \int {|\Pi({\bf x},t;\tau)|^2\over \mu}~dt~d^3{\bf x} \nonumber \\
  \mathbb{V}[\Psi] & = -  S_{\textrm{nr}}[\Psi({\bf x},t;\tau)]
  \label{eq:ham-T-V-2}
\end{align}
The symplectic dynamical equations of the Schr\"odinger field
$\Psi({\bf x},t;\tau)$ read then
\begin{align}
\partial_{\tau}{\Psi}({\bf x},t;\tau) 
&= \frac{\delta \mathbb{H}_{\text{nr}}[\Pi,\Psi] }{\delta \Pi^*({\bf x},t;\tau)} \; , 
\label{rep1}
\\
\partial_{\tau}{\Pi}({\bf x},t;\tau) & 
= - \frac{\delta \mathbb{H}_{\text{nr}}[\Pi,\Psi] }{\delta \Psi^*({\bf x},t;\tau)} \; .
\label{rep2}
\end{align} 
For the non-relativistic limit of the field-theory,
Eq.~\eqref{eq:lagrangian}, they take the explicit expressions
\begin{eqnarray}
\partial_{\tau}{\Psi}({\bf x},t;\tau) 
&=& {\Pi({\bf x},t;\tau)\over \mu} \; , 
\label{eq:nonrel-dyn-0}
\\
\partial_{\tau}{\Pi}({\bf x},t;\tau) &=& 
\left( i \hbar \partial_t + 
\frac{\hbar^2}{2m} \partial_{\bf x}^2 - g |\Psi|^2 \right) \Psi({\bf x},t;\tau)  
\; .  \nonumber \\
\label{eq:nonrel-dyn}
\end{eqnarray}

Before proceeding further in illustrating more details about the
outcomes of the non-relativistic limit of the symplectic quantization
scheme, we want to point out a crucial comment about its physical
interpretation. The deterministic dynamics in the intrinsic time
$\tau$, engendered by
Eqn.s~\eqref{eq:nonrel-dyn-0},\eqref{eq:nonrel-dyn}, makes the action
$S_{\textrm{nr}}[\Psi]$ play the role of an effective potential energy
$-\mathbb{V}[\Psi]$, which is, accordingly, a fluctuating quantity. In
the extended phase-space of the Schr\"odinger field $\Psi({\bf
  x},t;\tau)$ and of its conjugated momentum $\Pi({\bf x},t;\tau)$
these fluctuations are of quantum origin.

\section{Connection with standard non-relativistic QFT}
\label{sec:2-path-integral}

In this section we want to elaborate about the relation between the
non-relativistic symplectic quantization scheme introduced in the
previous section and the standard formulation of non-relativistic
field theory (QFT), based on the path integral formalism. The key
point is the assumption that the dynamics engendered by
Eqn.s~\eqref{rep1},\eqref{rep2} is ergodic in a generalized
microcanonical ensemble. \ggcol{Note that despite one would like to
  have at disposal a proof rather than an assumption, ergodicity is
  quite difficult to be established on a rigorous ground. Anyway,
  there is a general consensus that such an assumption holds for
  models of infinite-dimensional variables, like field theories.} Said
differently, ergodicity amounts to assume the existence of a uniform
probability measure $\mathcal{P}_\mA(\Pi,\Psi)$ in the
infinite-dimensional phase-space of fields and conjugate momenta
reading as:
\begin{equation}
{\cal P}_\mA[\Pi,\Psi] = {\delta(\mathbb{H}_{\textrm{nr}}[\Pi,\Psi] - {\cal A}) \over 
  \Omega({\cal A})},
\label{eq:micro-prob-nr}
\end{equation}
where the normalization factor is a generalized
microcanonical partition function fixing to ${\cal A}$ the value of the generalized 
non-relativistic Hamiltonian $\mathbb{H}_{\textrm{nr}}$ (which has
the actual physical dimensions of an action) :
\begin{equation}
\Omega({\cal A}) = \int {\mathcal D}[\Pi] {\mathcal D}[\Psi] 
\delta(\mathbb{H}_{\textrm{nr}}[\Pi,\Psi] - {\cal A}) \; ,  
\label{parvol}
\end{equation}
The ergodic hypothesis amounts then to assume that the dynamical
average along the trajectories of the symplectic quantization dynamics
do correspond to ensemble averages with respect to the probability
measure ${\cal P}_\mA[\Pi,\Psi]$ in Eq.~\eqref{eq:micro-prob-nr}:
\begin{align}
\lim_{\tau \to +\infty} \frac{1}{\tau} & \int_0^{\tau} 
    {\cal O}(\Psi_\mA({\bf x},t;\tau')) \, d\tau' = \nonumber \\
    = & \int {\mathcal D}[\Pi] {\mathcal D}[\Psi]~{\cal O}(\Psi)~{\cal P}_\mA[\Pi,\Psi],
\label{eq:erg-hyp}
\end{align}
where $\Psi_\mA({\bf x},t;\tau')$ denotes a solution of symplectic
dynamics equations relative to a generic initial condition, such that
$\mathbb{H}_{\textrm{nr}}[\Pi({\bf x},t;0),\Psi({\bf x},t;0)] =
\mathcal{A}$.  In Eqn.s~\eqref{parvol} and \eqref{eq:erg-hyp} the
symbols ${\mathcal D}[\Pi]={\mathcal D}[\Pi({\bf x},t)]$ and
${\mathcal D}[\Psi]={\mathcal D}[\Psi({\bf x},t)]$ denote functional
integration over the fields $\Pi({\bf x},t)$ and $\Psi({\bf x},t)$,
respectively. According to Eq.~\eqref{eq:erg-hyp} the dynamical average
can be replaced by the (microcanonical) ensemble average: for this
reason the fields and their conjugated momenta in the integral on the
r.h.s.  do not keep any dependence on $\tau$. This is in complete
analogy to what is assumed in classical statistical mechanics, where
the dynamics is replaced by (functional) integration over equilibrium
configurations. In this case the partition function in
Eq.~\eqref{parvol} turns out to be just the sum over all classical
configuration of the field and conjugated momenta, compatible with a
given value of the generalized action.\\

{\it It is worth stressing that the ergodic hypothesis is the core of
  the connection between symplectic quantization and ordinary quantum
  field theory also in the non-relativistic limit discussed here.}\\

Then, since in the present framework the action ${\cal A}$ is the
symplectic analog of the microcanonical internal energy, it is natural
to introduce the dimensionless symplectic analog of the familiar
microcanonical entropy as
\begin{equation} 
S_{\text{sym}}({\cal A}) = \ln\Omega({\cal A}) \; . 
\end{equation}
In this way, the reduced Planck constant $\hbar$ fixes the constraint 
\begin{equation}
{1\over \hbar} = {\partial S_{\text{sym}}({\cal A})\over  \partial{\cal A}} \; ,  
\label{gggod}
\end{equation}
or, equivalently, 
\begin{equation}
{1\over \hbar} = {1\over \Omega({\cal A})}
{\partial \Omega({\cal A})\over \partial {\cal A}} \; ,  
\label{gggod1}
\end{equation}
which relates the extensive quantity ${\cal A}$ to the intensive
quantity $\hbar$. As previously discussed, in the symplectic
quantization $\hbar$ plays the role of the intensive thermal energy
$k_B T$. \\

At this point our purpose is to draw the precise connection between
the microcanonical representation of symplectic quantization and the
most familiar functional approach to quantum field theory, i.e. the
path-integral representation.  As we are going to show this bridge can
be established by performing a suitable change of equilibrium
statistical ensemble, which, in this case, does not amount to a
standard Legendre transform. Actually, in the {\it microcanonical}
representation of the symplectic quantization scheme the value of the
action over the phase-space is fixed and one would like to transform
this formalism to a sort of {\it canonical} representation fixing {\it
  on average} the action per unit volume. In equilibrium statistical
mechanics the analogous procedure yields a canonical representation
based on a Gibbs-Boltzmann weight.  Since in symplectic quantization
scheme (also in the present non-relativistic case) the functional with
respect to which we would like to consider the integral transform is
an action, i.e. a "non-positive" defined quantity, the only
transformation which can be performed is the Fourier one.
Accordingly, the Fourier transform of the partition volume in
Eq.~\eqref{parvol} reads
\begin{align}
{\cal Z}(z) & = {1\over \sqrt{2\pi}} \int_{-\infty}^{+\infty} 
d\mA~e^{-iz\mA} \, \Omega({\cal A}) \nonumber \\
& = \int {\mathcal D}[\Pi] {\mathcal D}[\Psi] e^{-iz \mathbb{H}_{\textrm{nr}}[\Pi,\Psi]} \; ,  
\end{align}
where $z$ is the variable conjugated to $\mA$ and has therefore
physical dimensions of an inverse action, $[z]=[\mA^{-1}]$. By then
fixing $z=1/\hbar$ and integrating over the quadratic dependence on
conjugated momenta one obtains
\begin{eqnarray}
\label{part_fun}
{\cal Z}(\hbar) &=& \int {\mathcal D}[\Pi] {\mathcal D}[\Psi] \, 
e^{-{i\over \hbar} \mathbb{H}_{\textrm{nr}}[\Pi,\Psi]} 
\nonumber 
\\
&=& {\cal N}(\hbar) \int {\mathcal D}[\Psi] \, e^{{i\over \hbar} S_{\textrm{nr}}[\Psi]}, 
\end{eqnarray}
which is the standard Feynman path integral representation of the
non-relativistic QFT of the Schr\"odinger field $\Psi({\bf x},t)$,
where ${\cal N}(\hbar)$ is a suitable normalization factor.  We can
conclude that within the symplectic quantization approach the Feynman
path-integral is obtained, on the basis of the ergodic hypothesis
mentioned in this paragraph, simply as the Fourier transform of a
generalized microcanonical partition function based on the
conservation of the symplectic action. Only on a {\it formal} ground,
we can therefore think at Eq.~\eqref{part_fun} as {\it a sort of}
partition function and its complex integrand as {\it a sort of}
probability measure $\mP_\hbar[\Pi,\Psi]$ which, provided the original
ergodic hypothesis in Eq.~\eqref{eq:erg-hyp} holds true, conserves
{\it on average} the symplectic action, with $\mP_\hbar[\Pi,\Psi]$
reading as
\begin{align}
\mP_\hbar[\Pi,\Psi] = \frac{1}{\mZ(\hbar)} e^{-\frac{i}{\hbar} \mathbb{H}_{\text{nr}}[\Pi,\Psi]}.
\end{align}
More precisely, we can write
\begin{align}
\lim_{\tau \to +\infty} \frac{1}{\tau} & \int_0^{\tau} 
    {\cal O}(\Psi_\mA({\bf x},t;\tau')) \, d\tau' = \nonumber \\
    = & \int {\mathcal D}[\Pi] {\mathcal D}[\Psi]~{\cal O}(\Psi)~{\cal P}_\hbar[\Pi,\Psi] \nonumber \\
    = & \frac{1}{\mZ(\hbar)} \int {\mathcal D}[\Pi] {\mathcal D}[\Psi]~{\cal O}(\Psi)~e^{-\frac{i}{\hbar} \mathbb{H}_{\text{nr}}[\Pi,\Psi]} \nonumber \\
    = & \frac{\int {\mathcal D}[\Psi]~{\cal O}(\Psi)~e^{\frac{i}{\hbar} S_{\text{nr}}[\Psi]}}{\int {\mathcal D}[\Psi]~e^{\frac{i}{\hbar} S_{\text{nr}}[\Psi]}}
\label{eq:erg-hyp-hbar}
\end{align}
which makes explicit the relation between symplectic quantization
dynamical averages and the expectation values in ordinary quantum
field theory.

Note that, as for Eq.~\eqref{eq:Ham-symp-rel} and Eq.~\eqref{eq:nr-H},
the integration over the conjugated momenta in the last line of
Eq.~\eqref{eq:erg-hyp-hbar} is straightfoward only if the generalized
Hamiltonian corresponding to the symplectic action is {\it
  separable}. Additional computational difficulties occur when one has
to deal with models exhibiting symmetry properties whose symplectic
quantization yields a non-separable generalized Hamiltonian of the
form
\begin{align}
A[\Pi,\Psi] = \mathbb{K}[\Pi,\Psi] + \mathbb{V}[\Phi] \, ,
\end{align}
as is it happens, for instance, in non-abelian gauge
theories~\cite{DFF83} and, most notably, in gravity~\cite{DFF83,G21}.\\ 

We conclude this section by discussing the physical implications of
the quantization constraint in Eq.~\eqref{gggod}, which draws a formal
correspondence between $\hbar$, in the symplectic quantization
approach, and the microcanonical definition of temperature $T$ in
ordinary statistical mechanics, $T^{-1} = \partial S(E)/\partial E$,
where $S$ is the microcanonical entropy and $E$ is the total conserved
energy. In order to do that we take inspiration from the Parisi-Wu
stochastic quantization approach, where the quantum fluctuations of
the fields are associated to a stochastic dynamics, which mimicks the
contact with a fictitious thermal bath at temperature
$\hbar$. Therefore, the main idea is that we need to set the
conditions for the deterministic dynamics of Eq.~\eqref{eq:nonrel-dyn}
as if it would be performed in contact with a thermostat at
temperature $\hbar$. This task can be achieved by assuming that a sort
of equipartition theorem applies in the symplectic quantization
scheme, as for an ergodic Hamiltonian dynamics of a classical system
at temperature $T$. In fact, in a classical Hamiltonian system with total energy $E$, made on N degrees
of freedom, whose kinetic energy is denoted by $K$, the equipartiton
of the energy amounts to assume that
\begin{align}
\label{equipa}
k_B T = \frac{E}{N} = \frac{2\langle K \rangle}{N} \, ,
\end{align}
where
\begin{align}
\langle K \rangle = \lim_{\Delta t \rightarrow\infty} \frac{1}{\Delta t} \int_{t_0}^{t_0+\Delta t} ds~K[\Pi(s)],
\end{align}
Accordingly, the angle brackets $\langle \bullet \rangle$ are a
shorthand notation for a time-average, which, consistently with the
ergodic hypothesis, is assumed to be independent of the initial
conditions. We can export this procedure in the framework of the
symplectic quantization scheme by introducing the following
correlation function for generalized momenta
\begin{align}
  & \langle \Pi^*({\bf x},t) \Pi({\bf y},t') \rangle = \nonumber \\
  & \lim_{\Delta \tau\rightarrow \infty} \frac{1}{\Delta\tau} \int_{\tau_0}^{\tau_0+\Delta \tau} ds~\Pi^*({\bf x},t;s) \Pi({\bf y},t';s),
\end{align}
where $\Pi^*({\bf x},t;s)$ and $\Pi({\bf y},t';s)$ denote solutions of
the Hamiltonian equations, Eq.~\eqref{eq:nonrel-dyn}, and $\tau_0$
indicates a time scale over which the system has already reached
stationarity. The analogous of Eq.~\eqref{equipa} reads
\begin{align}
\frac{1}{\mu}\langle \Pi^*({\bf x},t) \Pi({\bf y},t') \rangle = \frac{\hbar}{2}~\delta(t-t')~\delta({\bf x}-{\bf y}),
\label{eq:corr-pi-nonrel-real}
\end{align}
where $\mu$ is a suitable dimensional constant. By
Fourier-transforming this equation one obtains the {\sl quantization}
condition
\begin{align}
  \frac{1}{\mu}\langle \Pi^*({\bf k},\omega) \Pi({\bf k},\omega) \rangle = \frac{\hbar}{2}.
  \label{eq:h-cond-nonrel}
\end{align}
Note that for a lattice version of the theory, the condition in
Eq.~(\ref{eq:corr-pi-nonrel-real}) should be replaced with
\begin{align}
\frac{1}{\mu}\langle \Pi^*({\bf x}_i,t_k) \Pi({\bf x}_j,t'_l) \rangle = \frac{\hbar}{2}~\frac{1}{b a^3}~\delta_{kl}~\delta_{ij},
\label{eq:corr-pi-nonrel-real-lattice}
\end{align}
where $a$ is the lattice spacing along spatial directions and $b$ is
the lattice spacing along time direction.  Again, by
Fourier-transforming this equation one obtains
Eq.~\eqref{eq:h-cond-nonrel}, where the continuum variables ${\bf k}$
and $\omega$ are replaced by their discretized version. These
quantization conditions point out the special role played by $\hbar$
in the symplectic quantization scheme. In any practical computation by
numerical methods of the deterministic dynamics in
Eq.~\eqref{eq:nonrel-dyn} one has to consider the class of initial
conditions yielding a stationary evolution, compatible with such
quantization constraint.  More details about the possible recipes for
identifying the right class of initial conditions for the numerical
simulaton of the dynamics will be reported in a forthcoming paper by
one of the authors~\cite{GG23}.

\section{Coarse-graining: from Quantum Field Theory to 
Quantum Mechanics}
\label{sec:3}

In this section we illustrate how to obtain a non-relativistic theory
for a quantum wave-function $ \psi({\bf x},t)$, by applying a suitable
coarse graining procedure to the Schr\"odinger field $ \Psi({\bf x},t;
\tau)$.  We limit our considerations to local fields and
single-particle wavefunctions, disregarding the case of many-body
wavefunctions, since they are non-local objects unsuitable for
describing in the same framework non-relativistic and relativistic
systems, {\sl a fortiori} in the presence of interactions.

We consider the paradigmatic case of a
self-interacting quantum scalar field which is also in interaction
with a {\it classical} electromagnetic field. Starting from the
Lagrangian in Eq.~\eqref{eq:lagrangian}, one can add the interaction
with the electromagnetic classical field $A^\mu({\bf x},t) = A^\mu(x)
= (\Phi(x),{\bf A}(x))$, with $\Phi({\bf x},t)$ the scalar potential
and ${\bf A}({\bf x},t)$ the vector potential, in a way which is
both covariant and gauge-invariant, by introducing the covariant
derivatives:
\begin{align}
{\partial_t}&\to {\partial_t} + i {q\over \hbar} \Phi \\
{\boldsymbol \nabla} &\to {\boldsymbol \nabla} - i {q\over \hbar} {\bf A}.
\end{align}
The classical nature of the electromagnetic field excludes any dependence of 
$A^\mu({\bf x},t)$ and $\Phi({\bf x},t)$ on the intrinsic time of
quantum fluctuations $\tau$. Since we assume that the elecromagnetic field is
a static external field, we can also neglect
 ``propagative'' terms such as $F^{\mu\nu}F_{\mu\nu}$.
Thereofere, the total Lagrangian, including the contributions of 
the external fields, reads 
\begin{align}
\mL_{\text{tot}} &= \mL(\varphi,\partial_{\tau}{\varphi})  + 
{q^2\over 2mc^2} |\varphi|^2 \Phi^2 
- {iq\hbar\over 2mc^2} (\varphi^*\partial_t\varphi - 
\varphi\partial_t\varphi^*) \Phi 
\nonumber 
\\
& - {q^2\over 2m} |\varphi|^2 {\bf A}^2 + {iq\hbar\over 2m} 
(\varphi^* {\boldsymbol\nabla}\varphi 
- \varphi{\boldsymbol\nabla}\varphi^*)\cdot {\bf A} 
\label{eq:kari1}
\end{align}
In order to consider the non-relativistic limit, we rewrite
$\varphi({\bf x},t;\tau)$ in terms of a high-frequency component
$e^{-i m c^2 t/\hbar}$ and a {\it slow} component $\Psi({\bf
  x},t;\tau)$ according to the relation in Eq.~\eqref{eq:slow-field}.
This replacement amounts to a sort of adiabatic elimination procedure, yielding
the following Lagrangian of the field
$\Psi(x,\tau)$
\begin{align}
&\mL_{\text{tot}} = \mL_{\textrm{nr}}[\Psi] + {q^2\over 2mc^2} |\Psi|^2 \Phi^2 
- q |\Psi|^2 \Phi \nonumber \\
&- {q^2\over 2m} |\Psi|^2 {\bf A}^2 + {iq\hbar\over 2m} 
(\Psi^* {\boldsymbol\nabla}\Psi - 
\Psi{\boldsymbol\nabla}\Psi^*)\cdot {\bf A} \, ,
\end{align}
where the non-relativistic Lagrangian $\mL_{\textrm{nr}}[\Psi]$ is the
one define in Eq.~\eqref{eq:action2}. Without loss of generality we can assume 
a null vector
potential, i.e. ${\bf A}=0$. By replacing $1/c^2$ with
$\epsilon_0\mu_0$  we can rewrite the complete Lagrangian 
in the form
\begin{align}
  &\mL_{\text{tot}}[\Psi({\bf x},t;\tau)] = \nonumber \\
& {i \hbar\over 2} \left( \Psi^*({\bf x},t;\tau) \partial_t 
\Psi({\bf x},t;\tau) - \Psi({\bf x},t;\tau) \partial_t \Psi^*({\bf x},t;\tau) 
\right) \nonumber \\
& - \frac{\hbar^2}{2m} |\partial_{\bf x} \Psi({\bf x},t;\tau)|^2 -
\frac{g}{2} |\Psi({\bf x},t;\tau)|^4 \nonumber \\
& + \left( \frac{\epsilon_0\mu_0 q^2}{2m} \Phi^2({\bf x},t) - 
q \Phi({\bf x},t) \right)~|\Psi({\bf x},t;\tau)|^2 \, ,
\label{eq:lagrangian-EM-nr}
\end{align}
where we have made explicit the 
dependence of the quantum fields on $\tau$, as opposed to the classical
fields, which are assumed independent of it.\\

We can now introduce the coarse-graining argument which allows us to
deduce quantum mechanics from quantum field theory. Let us recall that
within the symplectic quantization approach we have to deal with a
generalized Hamiltonian, that we have also called {\it symplectic
  action}, because its physical dimensions are those of an action.
This generalized Hamiltonian is related to the Lagrangian in
Eq.~\eqref{eq:lagrangian-EM-nr} and to the generalized conjugated
momenta appearing in the equation of motion:
\begin{align}
& \partial_{\tau}{\Pi}({\bf x},t;\tau) = 
\left( i \hbar \partial_t + 
\frac{\hbar^2}{2m} \partial_{\bf x}^2 - g |\Psi|^2 \right) 
\Psi({\bf x},t;\tau)  \nonumber \\
& + \left( \frac{\epsilon_0\mu_0 q^2}{2m} \Phi^2({\bf x},t) - 
q \Phi({\bf x},t) \right)\Psi({\bf x},t;\tau).
\label{almostS}
\end{align}
Since for physical reasons we have to assume that the generalized
Hamiltonian takes a finite value, also the ``potential energy''
contribution, i.e. the one depending on $\Psi({\bf x},t;\tau)$, and
the ``kinetic energy'' contribution, i.e. the one which depending on
$\Pi({\bf x},t;\tau)$, have to remain finite for any values of
$\tau$. Accordingly, the generalized momenta are bounded functions of
$\tau$.  Therefore, averaging with respect to the intrinsic time
$\tau$ both terms of Eq.~\eqref{almostS} we obtain that the l.h.s. of
this equation vanishes with $\tau \to +\infty$, namely
\begin{equation}
\lim_{\tau \to +\infty} \frac{1}{\tau} \int_0^{\tau} \, \frac{\partial 
\Pi({\bf x},t;\tau')}{\partial \tau'} \, d \tau' = \lim_{\tau \to +\infty} 
\frac{\Pi(\tau) - \Pi(0)}{\tau} \to 0
\end{equation}
For what concerns the quantum field  $\langle \Psi \rangle$, we assume that its intrinsic-time average
\begin{align}
\langle \Psi({\bf x},t;\tau) \rangle_\tau = 
\lim_{\tau \to +\infty} \frac{1}{\tau} \int_0^{\tau} \, 
\Psi({\bf x},t;\tau') \, d\tau'
\end{align}
{\it defines} the quantum mechanics wave-function
\begin{align}
\psi({\bf x},t) = \langle \Psi({\bf x},t;\tau) \rangle_\tau \, .
\end{align}
By assuming that the external field is such that  $q^2 \epsilon_0 \mu_0 \Phi^2/m \ll q \Phi$,
Eq.~\eqref{almostS} can be rewritten in terms of the wave-function as follows
\begin{align}
  & i\hbar \frac{\partial}{\partial t} \psi({\bf x},t) = \nonumber \\
  & - \frac{\hbar^2}{2m} {\bf \nabla}^2 \psi({\bf x},t) + q \psi({\bf x},t)
\Phi({\bf x},t) + g \langle |\Psi|^2 \Psi \rangle \, .
\label{almostS2}
\end{align}
The important information which
must be retained from Eq.~\eqref{almostS2} is that {\it in general}
the coarse-graining of the quantum fields over the fast dynamics over
$\tau$ does not yield a simple quantum-mechanical theory. The
average/coarse-graining of the non-linear interaction term with
respect to the sequence of quantum fluctuations controlled by $\tau$
does not yield an expression which can be written in terms of the
wavefunction $\psi({\bf x},t)$ only, because 
\begin{align}
\langle |\Psi|^2 \Psi \rangle ~\neq~ |\langle \Psi \rangle|^2 \langle 
\Psi \rangle. 
\end{align}
It is only in peculiar cases, related to quantum phase
transitions, that one finds situations
where the following replacement is correct
\begin{align}
\langle \hat{\Psi}^+({\bf x},t) \hat{\Psi}({\bf x},t) 
\hat{\Psi}({\bf x},t) \rangle = |\psi({\bf x},t)|^2 \psi({\bf x},t). 
\label{eq:coherent-phase}
\end{align}
where in Eq.~\eqref{eq:coherent-phase} the second quantization
formalism has been restored: the fluctuating field $\Psi({\bf
  x},t;\tau)$ has been replaced with the field operator
$\hat{\Psi}({\bf x},t)$ and $\langle \bullet \rangle$ denotes the
average over the Fock space rather than the intrinsic time average. A
replacement as the one of Eq.~\eqref{eq:coherent-phase} is possible
for instance when the average is made with respect to a coherent state
$|CS\rangle$, that is an eigenstate of the field operator
$\hat{\Psi}({\bf x},t)$, such that $\hat{\Psi}({\bf x},t)|CS\rangle =
\psi({\bf x},t) |CS\rangle$ (see, for instance, \cite{sala-book}).
However, we are not going to deal with these peculiar cases in this
paper.\\

The main conclusion is that for local fields in the presence of
self-interactions, in general the Schr\"odinger field $\Psi({\bf
  x},t;\tau)$ is intrinsically fluctuating and cannot be reduced to a
wave-function. Only when the effect of the external classical field is
dominant with respect to the self interaction, namely
\begin{align}
|| q \psi({\bf x},t) \Phi({\bf x},t) || \gg || g \langle |\Psi|^2 \Psi 
\rangle ||,
\end{align}
where $|| \cdot ||$ denotes an $L_2$ norm in the appropriate Hilbert
space, or when the self interaction is set to zero ($g=0$), the
coarse-graining with respect to the intrinsic time $\tau$ allows one to map the
fluctuating quantum field to a wave-function, uniquely determined in
each point of space-time as the solution of the Schr\"odinger equation:
\begin{align}
i\hbar \frac{\partial}{\partial t} \psi({\bf x},t) = 
\left[ -\frac{\hbar^2}{2m} {\bf \nabla}^2 + q \Phi({\bf x},t) \right] 
\psi({\bf x},t)
\end{align}

In summary, the above derivation is a well-defined protocol for
obtaining a wave function $\psi({\bf r},t)$ (with $|\psi({\bf
  x},t)|^2$ representing a probability density in space-time) from a
fluctuating quantum field $\Psi({\bf r},t;\tau)$ by a sort of
adiabatic elimination procedure, combined with suitable assumptions
about the role of self-interacting contributions in the original field
theory.

\section{Conclusions and Perspectives}
\label{sec:3a}
 
In this paper we have explained how the symplectic quantization
approach to quantum field theory introduced in~\cite{GL21,G21} can be
extended to the non-relativistic field theory of a Schr\"odinger field
and how, under a very reasonable {\it ergodic hypothesis} for the
deterministic dynamics of this new approach, it is possible to readily
obtain a connection with the standard functional approach to QFT based
on the Feynman path-integral formalism. Moreover, we have illustrated
how a {\it coarse-graining} procedure applied to the symplectic
quantization scheme yields ordinary quantum mechanics from a theory of
fluctuating quantum fields. This procedure is effective only for a
free quantum field or in the presence of an external classical field
much stronger than the self-interaction term of the quantum field.  In
Appendix~\ref{app:A} we discuss how the symplectic quantization
formalism applies to a non-relativistic particle in a potential
well. Last but not least we have pointed out how the symplectic
quantization dynamics, which flows along the new intrinsic time $\tau$
of quantum fluctuations, can provide new numerical protocols for
sampling quantum fluctuations of fields in real coordinate time $t$
also in the non-relativistic case. Further investigations should
certainly encompass numerical tests of the symplectic quantization
scheme for non-relativistic quantum fields (a first numerical study of
relativistic fields can be found in~\cite{GG23}) and the investigation
of the interplay between this formalism and alternative functional
approaches to study quantum fluctuations of non-relativistic fields
(in particular bosonic ones) already present in the literature,
e.g. the so-called Truncated Wigner Approximation~\cite{SLC02} or the
Stochastic Gross-Pitaevski equation~\cite{S99,CP08}.

\begin{acknowledgments}
We warmly thank F. Bigazzi, P. Di Cintio, S. Franz, D. Seminara and
S. Wimberger for useful discussions.  G.G. is partially supported by
the project MIUR-PRIN2022 {\it ``Emergent Dynamical Patterns of
  Disordered Systems with Applications to Natural Communities''} code
2022WPHMXK and acknowledges the Physics Department of Sapienza,
University of Rome, and the Physics and Astronomy Department ``Galileo
Galilei'', University of Padova, for kind hospitality during some
stages along the preparation of this manuscript. \\ R.L. acknowledges
partial support from project MIUR-PRIN2017 \emph{Coarse-grained
  description for non-equilibrium systems and transport phenomena}
(CO-NEST) n. 201798CZL. \\ L.S. is partially supported by the European
Union-NextGenerationEU within the National Center for HPC, Big Data
and Quantum Computing [Project No. CN00000013, CN1 Spoke 10: Quantum
  Computing], by the BIRD Project \emph{Ultracold atoms in curved
  geometries} of the University of Padova, by ``Iniziativa Specifica
Quantum'' of Istituto Nazionale di Fisica Nucleare, by the European
Quantum Flagship Project "PASQuanS 2", and by the PRIN 2022 Project
\emph{Quantum Atomic Mixtures: Droplets, Topological Structures, and
  Vortices" of the Italian Ministry for University and
  Research}. L. S. also acknowledges the Project "Frontiere
Quantistiche" (Dipartimenti di Eccellenza) of the Italian Ministry for
Universities and Research.
\end{acknowledgments}

\appendix
\section{Non-relativistic particle in a 
one-dimensional external potential}
\label{app:A}

We complete our discussion on the non-relativistic limit of symplectic
quantization by considering the one-dimensional non-relativistic
problem of a particle in an external potential. The classical
lagrangian of this non-relativistic system can be written as
\begin{equation}
L(q,{\dot q}) = {m\over 2} {\dot q}^2 - V(q) \; , 
\end{equation}
where $q(t)$ denotes the coordinate of a particle of mass $m$ as a
function of coordinate time $t$ and ${\dot q}$ is the derivative of
the coordinate with respect to $t$, while $V(q)$ is the external
potential acting on the particle. The action functional is given by
\begin{equation}
S[q(t)] = \int_{t_0}^{t_1} L(q,{\dot q}) \, dt = \int_{t_0}^{t_1}
\left( {m\over 2} {\dot q}^2 - V(q) \right) \, dt \; ,
\end{equation}
where $t \in [t_I,t_F]$, i.e. with $t_I$ and $t_F$ the 
two limits of integration over the time $t$. 

As previously discussed, the prescription of symplectic quantization
is that any classical field, in this case the ``position field''
$q(t)$, is ``quantized'' by assuming for it a further dependence on
the intrinsic time $\tau$, so that for any given value of coordinate
time $t$ the field $q(t;\tau)$ is not fixed but fluctuates along the
symplectic dynamics.

According to what already done for fields in the previous sections, we
postulates the existence of a generalized ``Lagrangian''
\begin{eqnarray}
\mathbb{L}[q,\partial_{\tau}{q}] & = & 
\int_{t_I}^{t_F} {M\over 2} (\partial_{\tau}{q}(t;\tau))^2 \, 
dt + S[q(t;\tau)] 
\\
& = & \int_{t_I}^{t_F} \left( 
{M\over 2} (\partial_{\tau}{q})^2 + 
{m\over 2} {\dot q}^2 - V(q) \right) \, dt \; ,
\nonumber 
\end{eqnarray}
where $\partial_{\tau}$ denotes again the derivative with respect to
the intrinsic time $\tau$ and $M$ is an appropriate dimensional
constant. In this case
\begin{align}
[M] = \textrm{mass} \; . 
\end{align}

The symplectic momentum reads 
\begin{align}
\pi(t;\tau) &= \frac{\delta\mathbb{L}}{\delta \partial_{\tau}{q}(t;\tau)} = 
M \partial_{\tau}{q}(x;\tau) \nonumber \\
\label{pip0}
\end{align}
and the symplectic-action/generalized-Hamiltonian functional is 
\begin{eqnarray}
\mathbb{H}[\pi,q] &=& \int_{t_I}^{t_F} \pi(t;\tau) \partial_{\tau}{q}(t;\tau) \, dt
- \mathbb{L}[q,\partial_{\tau}{q}] 
\nonumber 
\\
&=& \int_{t_I}^{t_F} {\pi(t;\tau)^2\over 2M} \ dt - S[q,\partial_{\tau}{q}]  
\\
&=& \int_{t_I}^{t_F} \left( {\pi(t;\tau)^2\over 2M} - 
{m\over 2} {\dot q}(t;\tau)^2 + V(q(t;\tau)) \right) \, dt \; . 
\nonumber 
\label{amio}
\end{eqnarray}

The symplectic dynamical equations are given by 
\begin{align}
\partial_{\tau}{q}(t;\tau) &= \frac{\delta \mathbb{H}[\pi,q]}
{\delta \pi(t;\tau)} \; , 
\\
\partial_{\tau}{\pi}(t;\tau) &= - \frac{\delta \mathbb{H}[\pi,q]}
        {\delta q(t;\tau)} \; .
\end{align} 
Explicitly, we have 
\begin{align}
\partial_{\tau}{q}(t;\tau) &= {\pi(t;\tau)\over M} \; , 
\label{pip1}
\\
\partial_{\tau}{\pi}(t;\tau) &= - m {\ddot q}(t;\tau) 
- \partial_q V(q(t;\tau)) \; .
\label{pip2}
\end{align} 
Clearly, Eq. (\ref{pip1}) is equal to Eq. (\ref{pip0}). \\

As done in the discussion of quantum fields, we need to define the
quantization constraints also for the symplectic quantization approach
to ordinary quantum mechanics. Following the discussion above in the
main text, this is simply obtained by asking the following
for the dynamical average of momenta
\begin{align}
\frac{1}{2M} \langle \pi(t) \pi(t') \rangle = \frac{\hbar}{2}~\delta(t-t'),
\end{align}
which, in Fourier space, reads as
\begin{align}
  \frac{1}{2M} \langle \pi(\omega) \pi(-\omega) \rangle = \frac{\hbar}{2},
  \label{eq:quant-constr-mq-Fourier}
\end{align}
telling us that each frequency of the ``momentum field'' Fourier
components, $\pi(\omega)$, carries a {\it half} quantum of action, the
other half being carried by the ``position field''. Therefore, if one
is about doing a numerical simulation on a discrete and finite time
grid, where also frequencies are discretized, the quantization
constraint of Eq.~\eqref{eq:quant-constr-mq-Fourier} can be obtained
by choosing as initial condition
\begin{align}
  \frac{1}{2M} |\pi(\omega_i)|^2 &= \hbar \quad\quad \forall~i, \nonumber \\
  |q(\omega_i)|^2 &= 0 \quad\quad \forall~i.
  \label{eq:quant-constr-mq-Fourier-discr}
\end{align}


As previously discussed, given a generic observable 
${\cal O}$ which depends on the coordinate $q$, 
i.e. ${\cal O}(q)$, its symplectic time average is given by 
\begin{equation} 
\langle {\cal O}(q) \rangle = 
\lim_{\tau \to +\infty} \frac{1}{\tau} \int_0^{\tau} 
{\cal O}(q(t;\tau')) \, d\tau' \, . 
\label{giafatto}
\end{equation}

Again, under the assumption of symplectic ergodicity, 
this time average can be rewritten  
as the following micro-canonical statistical average

\begin{equation}
\langle {\cal O}(q) \rangle = \int {\mathcal D}[\pi] {\mathcal D}[q] \, 
{\cal P}[\pi,q] \, {\cal O}(q(t)) \; , 
\end{equation} 

where 

\begin{equation}
{\cal P}_\mA[\pi,q] = {\delta(\mathbb{H}[\pi,q] - \mA) \over 
\int {\mathcal D}[\pi] {\mathcal D}[q] \, \delta(\mathbb{H}[\pi,q]-\mA)} \;    
\end{equation}

where $\mA$ is a given fixed value of the action and
$\mathbb{H}[\pi,q]$ is given by Eq.~\eqref{amio} but clearly without
the dependence with respect to the intrinsic time $\tau$, i.e.

\begin{equation}
\mathbb{H}[\pi,q] = \int_{t_I}^{t_F} \left( {\pi(t)^2\over 2M} - 
{m\over 2} {\dot q}(t)^2 + V(q(t)) \right) \, dt \; . 
\end{equation}

As previously discussed, ${\cal A}$ is not arbitary. 
It is related to the reduced Planck constant $\hbar$ by the 
fundamental expression 

\begin{equation}
{1\over \hbar} = {\partial \over \partial {\cal A}} 
\ln\left( \int {\mathcal D}[\pi] {\mathcal D}[q] \, 
\delta(\mathbb{H}[\pi,q]-{\cal A}) \right) \; ,  
\end{equation}

or, equivalently, one can use Eq. (\ref{gggod1}), where in this specific 
case 

\begin{equation}
\Omega({\cal A}) = \int {\mathcal D}[\pi] {\mathcal D}[q] \, 
\delta\big(\mathbb{H}[\pi,q] -{\cal A}\big) \; .  
\end{equation}

\subsection{Microcanonical definition of $\hbar$}

As a toy model, we consider a very simple one-dimensional system, where 
$q(t)$ and $\pi(t)$ are substantially time independent, 
i.e. the kinetic energy $(m/2){\dot q}^2$ is negligible. 
Moreover, we assume that the potential energy is given by 
the quartic oscillator $V(q)=(\alpha/2) q^4$. It follows that 
\begin{eqnarray}
\Omega({\cal A}) &=& \int d\pi dq \, 
\delta\left( (t_F-t_I) \left( {\pi^2\over 2M} 
+ {\alpha\over 2} q^4 \right) -{\cal A}\right) 
\nonumber 
\\
&=& {1\over 2} \sqrt{2M\over (t_F-t_I)} 
\int dq {1\over \sqrt{{\cal A}-{\alpha\over 2}q^4(t_F-t_I)}} 
\nonumber
\\
&=& 1.748 \ \sqrt{2M{\cal A} \over \alpha^{1/4}(t_F-t_I)^{3/4}} \; . 
\end{eqnarray}
Consequently, we obtain 
\begin{equation}
{1\over \Omega({\cal A})} {\partial \Omega({\cal A})\over \partial {\cal A}} 
= {1\over 2{\cal A}}
\end{equation}
and from Eq. (\ref{gggod1}) we find 
\begin{equation}
{\cal A} = {\hbar \over 2} \; . 
\end{equation}

\subsection{Ehrenfest theorem} 

A fundamental relation which must be deduced from
Eqs. (\ref{pip1}) and (\ref{pip2}) 
in order to asses the consistency of the above
formalism with standard quantum mechanics is the Ehrenfest theorem. 
We start from the
Ehrenfest relations, which can be obtained without effort from the
Hamilton equations (\ref{pip1}) and (\ref{pip2}), simply by taking the
average over intrinsinc time, see Eq. (\ref{giafatto}). 
If the potential energy $V(q(t;\tau))$ is such that both 
$q(t;\tau)$ and $\pi(t;\tau)$ are bounded, we can write:
\begin{align}
\lim_{\tau\rightarrow\infty} \frac{1}{\tau}\int_{0}^{\tau} d\tau' 
\frac{\partial \pi(t;\tau')}{\partial\tau'} = \lim_{\tau\rightarrow\infty} 
\frac{\pi(t;\tau)-\pi(t;0)}{\tau} = 0,
\end{align}
namely 
\begin{align}
\langle \partial_\tau\pi(t) \rangle = 0 = m \langle 
{\ddot q}(t)\rangle  + \langle \partial_q V(q(t)) \rangle,
\end{align}
from which, by denotig as $m \dot{q}(t)$ the momentum conjugated to
$q(t)$ with respect to ordinary time, we have
\begin{align}
\frac{d}{dt} \langle p(t) \rangle = - \left\langle V'(q) \right\rangle,
\end{align}
while we have that, in the present context, the equation
\begin{align}
m \frac{d}{dt} \langle q(t) \rangle =  \langle p(t) \rangle, 
\end{align}
comes simply as a definition.\\


\begin{thebibliography}{99}

\bibitem{GL21} G. Gradenigo, R. Livi, {\it ``Symplectic quantization
  I: dynamics of quantum fluctuations in a relativistic field
  theory''}, Found. Phys. {\bf 51} (3), 66 (2021).

\bibitem{G21} G. Gradenigo, {\it ``Symplectic Quantization II:
  Dynamics of Space-Time Quantum Fluctuations and the Cosmological
  Constant''}, Found. Phys. {\bf 51} (3), 64 (2021).

\bibitem{LL82} L.D. Landau, E.M. Lifshitz, {\it ``Quantum
  Electrodynamics''}, Butterworth-Heinemann (1982).
  
\bibitem{PW81} G. Parisi and Y. Wu, \emph{``Perturbation theory
  without gauge fixing''}, Scientia Sinica {\bf 24}, 483-496 (1981).

\bibitem{CR82} D. J. E. Callaway, A. Rahman, {\it ``Lattice gauge theory
  in the microcanonical ensemble''}, Phys. Rev. D {\bf 28}, 1506 (1982).

\bibitem{DFF83} De Alfaro, Fubini, and Furlan, 
{\it ``On the Functional Formulation of Quantum Field Theory''}, 
Nuovo Cim. {\bf 74}, 365 (1983). 

\bibitem{R77} \ggcol{S. Ruffo, {\it ``A comparison between nonlocal models
  and quantum mechanics''}, Lettere al Nuovo Cimento {\bf 20}, 221
  (1977).}
  
\bibitem{G83} E. Gozzi, {\it ``The new functional approach 
to field theory by De Alfaro, Fubini and Furlani and its 
connection ot the Parisi-Wu stochastic quantization''}, 
Phys. Lett. B {\bf 130}, 183 (1983). 

\bibitem{DH88} P.H. Daamgard, H. H\"uffel, Stochastic 
Quantization, World Scientific (1988).

\bibitem{KN96} M. Kanenaga and M. Namiki, 
``On the Stochastic Quantization Method: Characteristics and Applications 
to Singular Systems'', Proceedings of the 
Fourth International Conference on Squeezed States and 
Uncertainty Relations, pp. 229-234 (1996). 

\bibitem{MM94} I. Montvay, G. M\"unster, Quantum Fields on a 
Lattice, Cambridge University Press, Cambridge (1994).

\bibitem{mattuck} R.D. Mattuck, 
A Guide to Feynman Diagrams in the Many-body Problem, 
Dover Publications (1992). 

\bibitem{GG23} M. Giachello, G. Gradenigo, {\it ``Symplectic
  quantization and the causal structure of space-time: numerical tests
  of a new approach to quantum field theory''}, in preparation.

\bibitem{sala-book} L. Salasnich, Quantum Physics of Light and Matter, 
Springer International Publishing, Cham (2019). 

\bibitem{SLC02} A. Sinatra, C. Lobo, Y. Castin, {\it ``The truncated
  Wigner method for Bose condensed gases: limits of validity and
  applications''}, J. Phys. B {\bf 35}, 3599-3631 (2002).
  
\bibitem{S99} H.T.C. Stoof, {\it ``Coherent Versus Incoherent Dynamics
  During Bose-Einstein Condensation in Atomic Gases''}, J. Low
  Temp. Phys. {\bf 114}, 1/2 (1999)

\bibitem{CP08} S.P. Cockburn, N.P. Proukakis, {\it ``The Stochastic
  Gross-Pitaevskii Equation and Some Applications''}, Laser Physics
  {\bf 19}, 558-570 (2008).

  
\end{thebibliography}

\end{document}